\documentstyle[11pt,aaspp4]{article}
\begin{document}

\title{Atomic Physics with the Goddard High Resolution Spectrograph on the 
$Hubble\,Space\,Telescope$. \\ {\rm V}. Oscillator Strengths for Neutral 
Carbon Lines below 1200~\AA \footnotemark[1]}
\author{S.R. Federman and J. Zsarg\'{o}}
\affil{Department of Physics and Astronomy, University of Toledo, Toledo, OH 
43606}

\footnotetext[1]{Based on observations obtained with the NASA/ESA $Hubble$ 
$Space$ $Telescope$ through the Space Telescope Science Institute, which is 
operated by the Association of Universities for Research in Astronomy, Inc.,
 under NASA contract NAS5-26555.}

\begin{abstract}
We analyzed high resolution spectra of interstellar C~{\small I} absorption
toward $\lambda$~Ori, 1~Sco, and $\delta$~Sco that were obtained with the 
Goddard High Resolution Spectrograph on the $Hubble\,Space\,Telescope$. 
Several multiplets were detected within the 
wavelength interval 1150 to 1200~\AA, 
where most C~{\small I} lines have ill-defined oscillator strength; multiplets 
at longer wavelengths with well-defined atomic parameters were also seen. We 
extracted accurate column densities and Doppler parameters from lines with 
precise laboratory-based $f$-values. These column densities and $b$-values 
were used to obtain a self-consistent set of $f$-values for all the observed 
C~{\small I} lines. For many of the lines with wavelength below 1200~\AA, the 
derived $f$-values differ appreciably from the values quoted in the compilation 
by Morton (1991). The present set of $f$-values extends and in some cases 
supersedes those given in Zsarg\'{o} et al. (1997), which were based on lower 
resolution data.
\end{abstract}


\section{Introduction}

Accurate oscillator strengths ($f$-values) are needed for spectroscopic 
studies in astronomy. For instance, they are required when extracting reliable 
abundances from interstellar absorption lines, 
when modeling opacities in stellar atmospheres, or when utilizing 
temperature and density diagnostics. While analyzing such spectra 
from space-borne, high-resolution UV spectrographs, one can encounter the 
problem that uncertainties associated with observational sources are less than 
those from atomic physics. This is especially true when the astronomical data 
have signal-to-noise ratios greater than 100 to 200. One can take advantage of 
the situation by refining $f$-values for lines giving discordant results.  
The basic premise involves obtaining column densities and Doppler parameters 
from lines where there is consensus on $f$-values and then using this 
information in refining other $f$-values.

Several recent studies based on interstellar spectra acquired with the Goddard 
High Resolution Spectrograph (GHRS) on the {\it Hubble Space Telescope} ($HST$) 
have adopted this methodology. Federman \& Cardelli (1995) provided new 
$f$-values for lines of S~{\small I}; many of their determinations were 
confirmed by subsequent theoretical (Tayal 1998) and experimental work 
(e.g., Biemont et al. 1998). Cardelli \& Savage (1995) analyzed Fe~{\small II} 
lines, and Zsarg\'{o}, Federman, \& Cardelli (1997) refined $f$-values for 
C~{\small I} lines with central wavelength below 1200~\AA\ as well as for 
some forbidden lines above this limit. Relative $f$-values were derived for 
singly-ionized nickel (Zsarg\'{o} \& Federman 1998) and singly-ionized cobalt 
(Mullman et al. 1998). The latter analysis was performed in parallel with 
laboratory measurements that placed astronomical oscillator strengths on an 
absolute scale. Laboratory measurements on Ni~{\small II} by Fedchak, Wiese, 
\& Lawler (2000) validated the relative $f$-values in our earlier work on 
singly-ionized nickel.

In the present paper, following the method outlined in Zsarg\'{o} 
et al. (1997), we improved upon their $f$-values for C~{\small I} lines with 
wavelength below 1200~\AA\ and expand the number of lines in their list. In 
Section~\ref{section:Obs} we briefly describe the astronomical measurements 
available for our analysis, and in Section~\ref{section:Model} we discuss how 
we obtained column densities and Doppler parameters for each 
absorbing component (\ref{subsection:ProfileSynthesis}) 
and how we adjusted $f$-values 
(\ref{subsection:revision}). Finally, we discuss our results in 
Section~\ref{section:Disc}.

\section{Measurements \label{section:Obs}}

We retrieved observations for $\lambda$~Ori, 1~Sco, and $\delta$~Sco from the 
$HST$ archive. Most of the measurements were acquired at medium resolution 
(MR) and a handful of them were taken at high resolution (HR). These spectra 
covered several C~{\small I} multiplets and forbidden lines between 1150~\AA\ 
and 1700~\AA.

The reduction was fairly straightforward, and only minor technical problems 
arose. The most serious concern was an error in the background correction on 
some of our preprocessed data. We detected this error on all of our HR spectra 
containing the $\lambda$1261 (UV9) multiplet of C~{\small I}. The flux level of 
strongly saturated lines, such as those of S~{\small II} and Si~{\small II}, 
should be zero at line center; however, in some cases we measured negative 
or small positive values, clearly indicating an erroneous 
background correction. 
We could then apply an additional correction to compensate for the effect. The 
continuum placement turned out to be difficult for some multiplets, due to 
blending of several lines or simply because the lines were saturated. For 
example, in the vicinity of multiplet $\lambda$1261 in MR spectra, 
Si~{\small II} and Fe~{\small II} lines seriously affected the continuum 
placement. Following line identification and continuum placement, the Doppler 
shift (relative to the laboratory wavelength) and equivalent width 
($W_{\lambda}$) of neutral carbon lines were measured. In some cases, only 
upper limits could be calculated based on the average 
full width at half maximum (FWHM) of the 
instrumental function and the root mean square (rms) average of the noise in 
the stellar continuum.

We now give a brief description of the available spectra for 
each of our targets and the data reduction done on them. We list the measured 
$W_{\lambda}$ values together with those calculated from synthetic profiles 
(see Section~\ref{subsection:ProfileSynthesis} for details on profile 
synthesis) and measured in earlier work.  We note that the agreement between 
our $W_{\lambda}$ values and those obtained with the $Copernicus$ satellite 
(Jenkins \& Shaya 1979; Jenkins, Jura, \& Loewenstein 1983) is very good.

\subsection{$\lambda$~Ori (HD36861)}

We retrieved eight observations of $\lambda$~Ori from the $HST$ archive (see 
Table~\ref{table:Obs}), two at high resolution and six at medium. Both HR 
spectra, z3di0204t and z3di0205t, were made with the Large Science Aperture 
(LSA) and so their resolution was slightly lower than usual 
($\approx$4 km~s$^{-1}$ instead of $\approx$3 km~s$^{-1}$). The only observable 
C~{\small I} lines in the range of HR spectra were those of the multiplets 
$\lambda\lambda$1157 (UV16), 1158 (UV15.01). There were two 
additional multiplets covered by the MR observations, $\lambda$1261 
and $\lambda$1329 (UV4). We could measure $W_{\lambda}$ 
for all resolved lines (see Table~\ref{table:lOriW} for measured and 
calculated values). The two resolved components are separated by $\approx$12 
km s$^{-1}$, as seen in Na~{\small I~D} spectra (Hobbs 1974). 

\subsection{1~Sco (HD141637)}

We retrieved eight measurements of 1~Sco from the $HST$ archive 
(Table~\ref{table:Obs}), two at high and six at medium resolution. The HR 
spectra contained three multiplets, $\lambda\lambda$1194 (UV9.02) 
and 1329 fully and $\lambda$1193 (UV11) partially. 
Multiplet $\lambda$1329 was very useful for our 
analysis, since it has very well established oscillator strengths. 
In MR spectra, we 
had measurements of all important multiplets with central wavelength between 
1200~\AA\ and 1700~\AA\ [$\lambda\lambda$1261, 1277 (UV7), 1280 (UV5), 
1329, 1561 (UV3), and 1657 (UV2)]. 
The measured and calculated $W_{\lambda}$ values, together with earlier 
measurements, are listed in Table~\ref{table:1ScoW}. The two components are 
separated by about 7 km s$^{-1}$ and correspond to the strongest components in 
Na~{\small I~D} (Welty, Hobbs, \& Kulkarni 1994).

\subsection{$\delta$~Sco (HD143275)}

We obtained 7 observations of $\delta$~Sco, as listed in Table~\ref{table:Obs}, 
from the $HST$ archive. Four of them were high resolution, while the rest were 
medium resolution. Two of the HR spectra, z3di0504t and z3di0505t, were made 
with the LSA; hence, their resolution was lower than usual. We could identify 
most members of the multiplets $\lambda\lambda$1157, 1158
in these observations. The other two HR observations (z34r020dt 
and z34r020et) were of very good quality. The Small Science Aperture (SSA) was 
utilized; therefore, they had the maximum possible resolution of the GHRS. Both 
covered multiplet $\lambda$1261 only, close to the upper wavelength end of the 
spectra, and the J=2 lines were out of the range of z34r020dt. 
We also had three MR spectra containing 
the full $\lambda$1261 multiplet. A combined MR spectrum was 
created to increase the signal-to-noise ratio. The measured and calculated 
$W_{\lambda}$ values, together with previous measurements, are given in 
Table~\ref{table:dScoW}. The 5 km s$^{-1}$ separation in components is seen in 
K~{\small I} also (Welty \& Hobbs 2000) and between the two CH$^+$ components 
found by Crane, Lambert, \& Sheffer (1995).

\section{Model \label{section:Model}}

\subsection{Profile Synthesis \label{subsection:ProfileSynthesis}}

Transitions belonging to the same fine structure level can be characterized by 
the same column density ($N$), Doppler or line-broadening parameter 
($b$-value), and Doppler shift. All the other 
parameters are independent of the line of sight (LOS), and are transition 
specific. In particular, line oscillator strengths, or $f$-values, are needed. 
Our task was to specify the number of absorbing 
components and the corresponding 
column densities and $b$-values for all observable fine stucture levels. 

The number of resolved components, which correspond to the strongest features 
seen in Na~{\small I}, K~{\small I}, and CH$^+$ spectra, was revealed by 
visual inspection. There might be, of course, some unresolved 
ones with separation between them less than the resolution. While slight 
differences in extracted column densities may result in our approach, the 
unresolved components could be treated as a single component for all of our 
modelling purposes.

We had two options to find the column densities and $b$-values. The first, a 
more traditional and simpler way, is the curve of growth (COG) analysis. While 
it is relatively easy to implement, it is less effective in finding the 
parameter values, since only the $W_{\lambda}$ is used and not the entire line 
profile. Also, the COG method is not suitable for analyzing blended lines. We 
used this method only to adjust oscillator strengths, whenever it was necessary 
or possible to do so [see Zsarg\'{o} et al. (1997) for details]. 

\begin{figure}[h]
\begin{center}
\plotfiddle{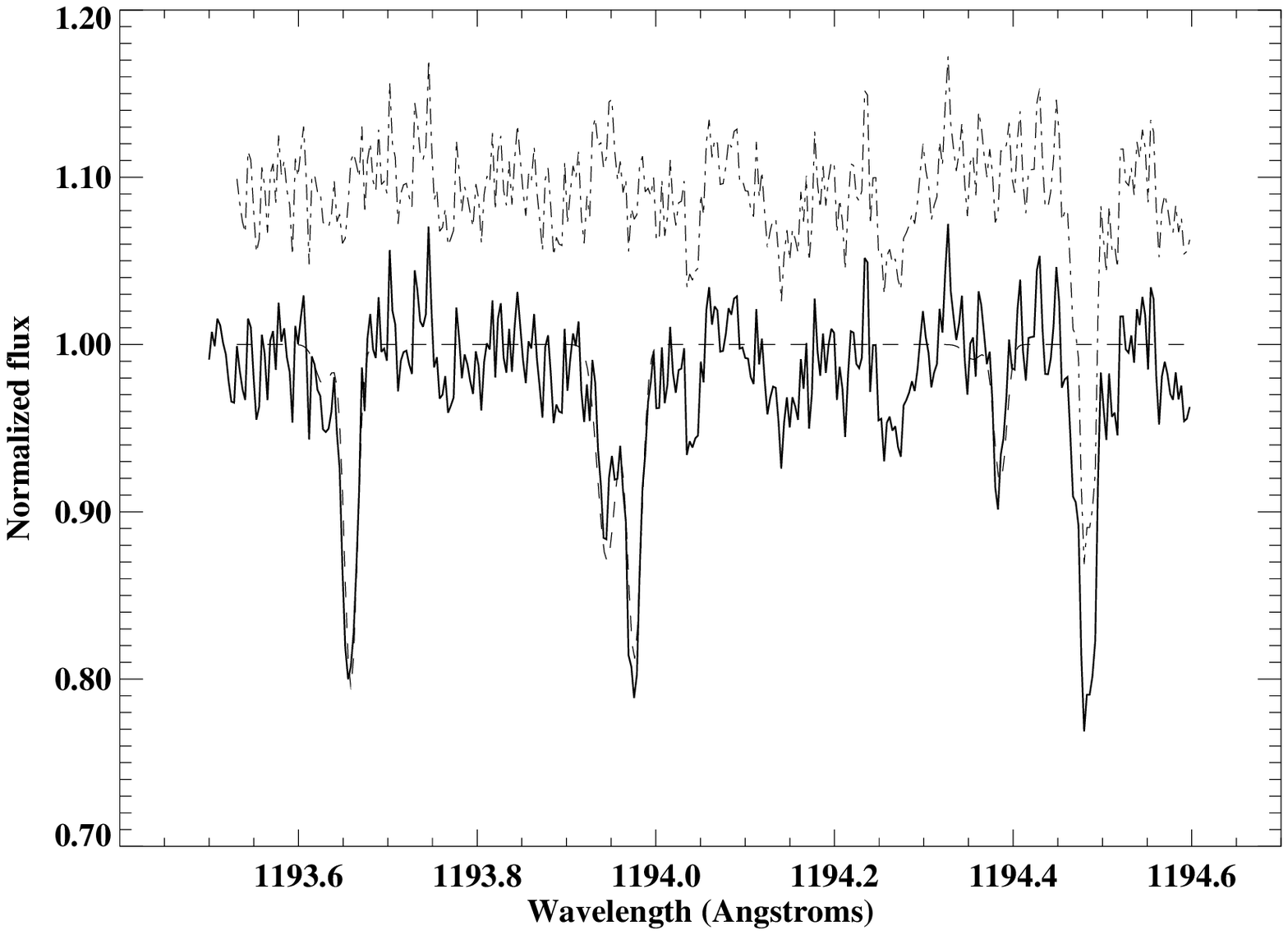}{5.5in}{0}{80}{80}{-225}{-15}
\figcaption{ High Resolution spectra of multiplet $\lambda$1194 toward 1~Sco. 
The solid and dashed lines are the observed and the calculated spectra, 
respectively, and the dash-dotted line shows the residuals (offset to +1.1). 
The Si {\small II} line at 1194.5 \AA\ was not synthesized.
\label{Fig. 1}}
\end{center}
\end{figure}

\begin{figure}[h]
\begin{center}
\plotfiddle{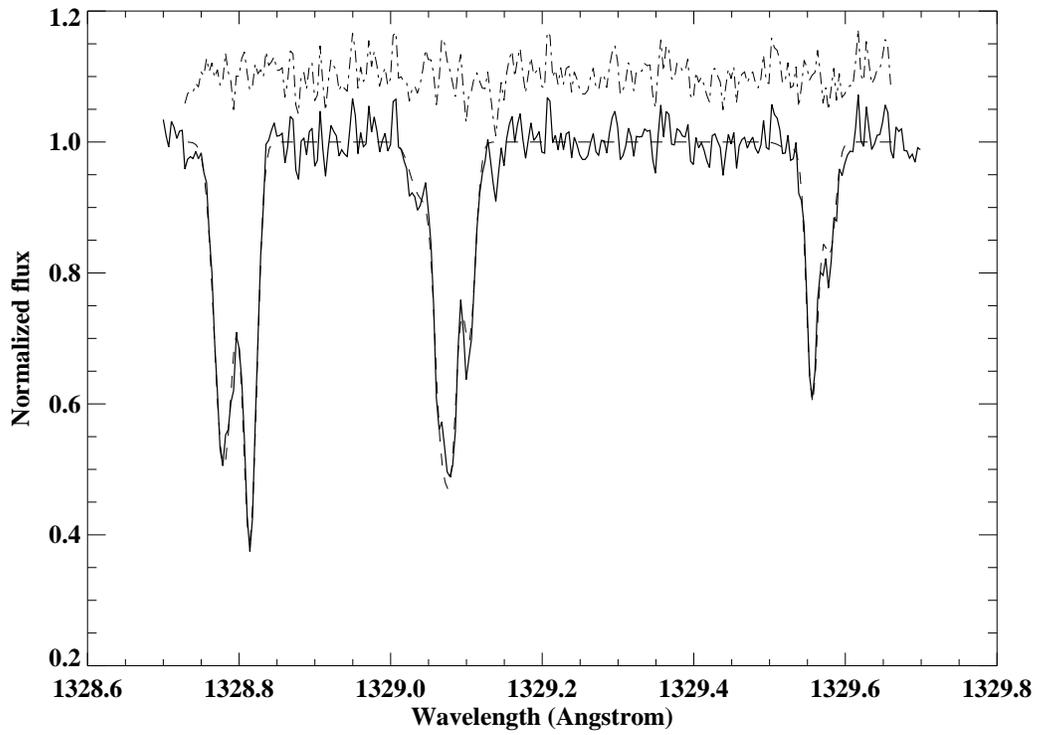}{5.5in}{0}{80}{80}{-225}{-15}
\figcaption{High resolution spectra of multiplet $\lambda$1329 toward 1 Sco. 
The definition of the line types is the same as in Fig~\ref{Fig. 1}. 
\label{Fig. 2}}
\end{center}
\end{figure}

We used the more complex and accurate method that matches the observed profiles 
with synthesized ones to find the LOS-specific parameters. We calculated the 
``observed'' fluxes for all spectra with a given set of column densities, 
component velocities, and $b$-values, and compared them to the corresponding 
observations. The set of initial parameters then was adjusted until a 
satisfactory fit could be achieved. 
The final sets of parameters appear in Table~\ref{table:fit} for our lines of 
sight. Here for simplicity, Doppler shifts are relative to the main velocity 
component which was set at $\sim$ 0 km s$^{-1}$ within IRAF, the Image 
Reduction and Analysis Facility from the National Optical Astronomy 
Observatories.  The slight differences in velocity 
between fine structure levels 
(usually less than $\pm$~0.5~km~s$^{-1}$) for the primary components are small 
compared to the spectral resolution of $\approx$~3.5~km~s$^{-1}$. 
The velocities were fixed for the secondary component in each case. 
Figures~\ref{Fig. 1} and \ref{Fig. 2} 
show the observed and synthesized HR spectra of multiplets 
$\lambda\lambda$1194 and 1329 for 1~Sco. We used the new oscillator strengths 
for multiplet $\lambda$1194 to produce the synthesized spectrum. Details of 
the profile synthesis will be presented in another paper (Zsarg\'{o}, 
Federman, \& Welty 2001).

\subsection{Revision of Neutral Carbon Oscillator Strengths 
\label{subsection:revision}}

The neutral carbon multiplets and forbidden lines that were used throughout 
our analysis covered the wavelength range 1150 to 1700~\AA. Every multiplet 
above 1200~\AA\ is thought to have well established $f$-values. Measurements 
and calculations have been performed, and a general consensus has been achieved 
about the values (see Morton 1991). However, for the 
multiplets with wavelengths 
below 1200~\AA\ and for some weak forbidden lines above it, only calculations 
are available. The lack of laboratory and precise 
astronomical measurements make 
these oscillator strengths less reliable.

The quality of GHRS spectra are generally very good, especially those with high 
spectral resolution. This provides a unique opportunity to fill in the gap of 
missing measurements. Consider the situation where HR spectra of two multiplets 
for a given LOS, one with well defined and the other with poorly defined 
oscillator strengths, are available. The first multiplet can be used to infer 
the LOS information, which is then the basis for refining the $f$-values for 
the second multiplet. Obviously, the more spectra of multiplets with 
well-defined oscillator strengths we possess, the more precise the adjustment 
will be.

We adjusted oscillator strengths with a curve of growth analysis. The 
adjustment of ill-defined $f$-values is accomplished by placing all 
($f \times \lambda$, $W_{\lambda} / \lambda$) points on the curve by a $\chi^2$ 
analysis. A special algorithm, described in detail by Zsarg\'{o} et al. 
(1997), was used to accomplish this goal. This algorithm allowed us to adjust 
the $f$-values either individually for each line or as a group for members of 
the same multiplet. It should be noted that there is danger in such an 
adjustment. Saturated lines lie on the flat part of the COG, and in such 
cases, a small uncertainty in $W_{\lambda}$ can result in a large error in 
$f$-value. Similar errors can occur if the $b$-value is ill defined. This 
parameter controls where the COG departs from the linear approximation; 
therefore, bad $b$-values can result in considerable errors in the values 
of the adjusted oscillator strengths. It is important to have very reasonable 
estimates for $b$-values and limit this approach for the weaker lines. While 
we could not quantify the effect of uncertainties in $b$-values, it is 
reassuring that we were able to get self-consistent results with the adjusted 
$f$-values toward multiple lines of sight (Zsarg\'{o} et al. 2001).

\section{Discussion \label{section:Disc}}

\subsection{Reanalysis of J=0 Data for $\beta^1$~Sco}

In the course of our work here, we found that Morton's (1991) $f$-value for 
$\lambda$1276 yielded more consistent results than the one given by us 
(Zsarg\'{o} et al. 1997). Closer inspection of our earlier analysis on 
$\lambda$1276 revealed that this line was on the flat portion of the curve of 
growth for $\rho$~Oph~A and $\chi$~Oph, and so its $f$-value could not be 
reliably determined from those data. This supports the inference made by 
Wannier et al. (1999) from HR spectra of C~{\small I} in the envelope of the 
molecular cloud B5.  [The other spin-forbidden transistions 
studied by Zsarg\'{o} et al. (1997) were weaker and lie on the linear portion 
of the COG.]

Unfortunately, our analysis of the J=0 lines in the spectrum of $\beta^1$~Sco 
was based in large measure on $\lambda$1276. If instead we use the multiplets 
$\lambda\lambda$1261, 1329, we obtain a column density of 
(1.0$\pm$0.1) $\times$ 10$^{14}$ cm$^{-2}$ and a $b$-value of 2.3 km~s$^{-1}$. 
These parameters lead to significant changes in $f$-values for J=0 lines below 
1200~\AA. The results presented in Table~\ref{table:Newfvalues} for J=0 are 
from the revised analysis. Since the J=1 and 2 lines originally were based on 
$\lambda\lambda$1261, 1329, no changes were made to their $f$-values.

\subsection{Comparison with Other Results}

We adjusted oscillator strengths for several lines and multiplets; they are 
displayed in Table~\ref{table:Newfvalues}. Both the new values and others are 
listed, together with the ratios of the $f$-values from Morton's (1991) 
compilation to our new values. The estimated 
uncertainties for the new oscillator strengths are given within 
parentheses and are based on the uncertainties in $W_{\lambda}$ and a 
10\% uncertainty in column density, taken in quadrature.  The uncertainty 
in column density is mainly the result of continuum 
placement. The agreement between the present results 
for J=0 and those of Zsargo et al. (1997), as corrected above, 
is generally very good. The 
differences involving lines in multiplet $\lambda$1157 may be caused by the 
use of MR spectra in our earlier effort. The comparison with the 
recommended values of Morton (1991) for the multiplets 
$\lambda\lambda$1193, 1194 is also very good, except for the line at 1193.6 
\AA; the correspondence for lines at shorter wavelengths is less 
satisfactory. Where significant differences are found, they appear to be the 
result of a breakdown in LS coupling rules for dipole-allowed transitions. 
The comparison with the theoretical 
work of Hibbert et al. (1993) shows that the results are very similar. The 
agreement with the $f$-values given by the compilation of Wiese, Fuhr, \& 
Deters (1996) is good for lines above 1158.5~\AA\ in large measure because 
they adopted the results of Hibbert et al. (1993). Below this wavelength our 
results suggest that LS coupling breaks down. For $\lambda$1193.6, where a 
factor-of-2 difference arises in the $f$-value of Welty et al. (1999), it 
is between ours and Morton's (1991). 

The ratios given in the last column are very useful in assessing whether LS 
coupling applies or not. Since Morton (1991) always assumed LS coupling [and 
so did Wiese et al. (1996) below 1180~\AA], any variation in the ratio within 
a multiplet indicates that this approximation does not apply. The results of 
using Cowan's (1981) atomic structure code (Zsarg\'{o} et al. 1997) can 
be used as a guide in describing the breakdown of LS coupling rules. As noted 
by Zsarg\'{o} et al. (1997), configuration interaction (CI) is stronger in 
$\lambda$1194 (5$s$ $^3P^{\rm o}_{0,1,2}$) 
than in $\lambda$1189 (UV14) and there is CI between $\lambda$1158 
(6$s$ $^3P^{\rm o}_{0,1,2}$) and $\lambda$1156 (UV19) for level J=1. 
Another prediction is CI between $\lambda$1158 and $\lambda$1156 for J=2, 
which we confirm now through observational data. We also see the predicted 
spin-orbit (SO) mixing between 2$s^2$ 2$p$ 6$s$ $^3P^{\rm o}_2$ and 
2$s^2$ 2$p$ 5$d$ $^3D^{\rm o}_2$ (of $\lambda$1157). No other strong effects 
are seen in $\lambda$1157, as predicted.

\subsection{Concluding Remarks}

In the course of a larger study involving C~{\small I} excitation (Zsargo et 
al. 2001), we found that lines below 1200~\AA\ 
gave inconsistent column densities and $b$-values. 
In the present work, we described refinements to the $f$-values 
for these lines so that a self-consistent analysis was possible. We readjusted 
the oscillator strengths for some lines in our 
earlier work (Zsargo et al. 1997) 
to correct for the (then) inappropriate choice of $\lambda$1276 as the basis 
of our analysis for the J=0 fine-structure level, and we find good agreement 
between results from the independent data sets.  For the few instances where 
differences are seen, the current $f$-values are preferred because they 
are based on HR spectra. We also supplemented our inventory of refined 
$f$-values with lines that were not available in our earlier work. 
The agreement 
between our and other recent compilations is generally good. Where 
disagreements arise, they mainly result from the assumption in other work that 
LS coupling applies 
for the multiplets $\lambda\lambda$1157 and 1158, an approximation not 
supported by our measurements and analysis.  Our extended set of 
$f$-values for C~{\small I} lines below 1200 \AA\ can be used with 
confidence in analyses of interstellar spectra acquired with the 
{\it Far Ultraviolet Spectroscopic Explorer}.

\acknowledgements

This research was supported by NASA grant NAG5-7754 and STScI grant 
AR-08352.01-A.

\newpage

\newpage

\newpage
\begin{table}[h]
\caption{Observations \label{table:Obs}}
\begin{tabular}{lccc} \hline
Archival code & Range        & Date of observation & Grating (aperture) \\
              &   (\AA)      &                     &                  \\ \hline
$\lambda$~Ori &              &                     &                  \\
z3di0204t     & 1157--1164   & 01/23/97            & Ech-A (LSA)      \\
z3di0205t     & 1158--1164   & 01/23/97            & Ech-A (LSA)      \\
z2c20306t     & 1226--1265   & 09/14/94            & G160M (SSA)      \\
z2c20307t     & 1230--1266   & 09/14/94            & G160M (SSA)      \\
z2c20308t     & 1231--1267   & 09/14/94            & G160M (SSA)      \\
z2520306t     & 1316--1352   & 02/19/94            & G160M (SSA)      \\
z2520307t     & 1326--1362   & 02/19/94            & G160M (SSA)      \\
z2520308t     & 1327--1363   & 02/19/94            & G160M (SSA)      \\
              &              &                     &                  \\
1~Sco         &              &                     &                  \\
z2zx020bp     & 1327--1334   & 03/14/96            & Ech-A (SSA)      \\
z2zx020fp     & 1188--1195   & 03/14/96            & Ech-A (SSA)      \\
z0zi0708m     & 1114--1150   & 07/26/92            & G160M (SSA)      \\
z0zi070bm     & 1229--1265   & 07/26/92            & G160M (SSA)      \\
z0zi070ct     & 1270--1306   & 07/26/92            & G160M (SSA)      \\
z0zi070dt     & 1325--1361   & 07/26/92            & G160M (SSA)      \\
z0zi070gt     & 1534--1569   & 07/26/92            & G160M (SSA)      \\
z0zi070it     & 1644--1679   & 07/26/92            & G160M (SSA)      \\ 
              &              &                     &                  \\ 
$\delta$~Sco  &              &                     &                  \\
z3di0504t     & 1157--1163   & 08/20/96            & ECH-A (LSA)      \\
z3di0505t     & 1158--1164   & 08/20/96            & ECH-A (LSA)      \\
z34r020dt     & 1254--1261   & 03/04/96            & ECH-A (SSA)      \\
z34r020et     & 1255--1262   & 03/04/96            & ECH-A (SSA)      \\
z2c20506t     & 1228--1265   & 04/23/94            & G160M (SSA)      \\
z2c20507t     & 1231--1267   & 04/23/94            & G160M (SSA)      \\
z2c20508t     & 1231--1267   & 04/23/94            & G160M (SSA)      \\ \hline
\end{tabular}
\end{table}

\newpage
\begin{table}[h]
\caption{$W_{\lambda}$ of neutral carbon lines toward $\lambda$~Ori 
\label{table:lOriW}}
\begin{tabular}{ccccccc} \hline
                &                &                    & &                  
&                       &                   \\
$\lambda$ & \multicolumn{2}{c}{Measured (m\AA)} & 
&  \multicolumn{2}{c}{Other Measurements}  & Calculated \\ 
\cline{2-3} \cline{5-6}
 (\AA)    & HR$^a$           & MR$^a$            & &     JS$^b$       
&    JJL$^c$       &    (m\AA)    \\
                &                &                    & &                  
&                       &                   \\ \hline \hline
J $=$ 0: &                  &                   & &                  
&                  &       \\
1157.910A$^d$ & 26.97 $\pm$ 0.26 & $\ldots$     & & $\ldots$         
& $\ldots$         & 28.09 \\
1157.910B$^d$ &  2.59 $\pm$ 0.30 & $\ldots$     & & $\ldots$         
& $\ldots$         & 4.25 \\
1158.324A & 15.28 $\pm$ 0.23 & $\ldots$         & & $\ldots$         
& $\ldots$         & 15.52 \\
1158.324B &  2.99 $\pm$ 0.35 & $\ldots$         & & $\ldots$         
& $\ldots$         & 1.52 \\
1260.736A & $\ldots$         & 35.44 $\pm$ 0.33 & & 37.0$^e$ $\pm$ 2.0 
& 38.0$^e$ $\pm$ 2.6 & 31.27 \\
1260.736B & $\ldots$         &  3.17 $\pm$ 0.24 & & $\ldots$         
& $\ldots$         & 4.88 \\
1328.833  & $\ldots$         & 48.23 $\pm$ 0.32 & & 52.0$^e$ $\pm$ 4.0 
& 51.0$^e$ $\pm$ 2.6 & 45.66 \\
J $=$ 1: &                  &                  & &                  
&                  &       \\
1157.770 &  5.82 $\pm$ 0.21 &  $\ldots$        & & $\ldots$         
& $\ldots$         & 5.58  \\
1158.035 &  8.95 $\pm$ 0.21 &  $\ldots$        & & $\ldots$         
& $\ldots$         & 9.56  \\
1158.130 &  4.42$^f$ $\pm$ 0.21 &  $\ldots$    & & $\ldots$         
& $\ldots$         & 4.28 \\
1158.674 &  0.85 $\pm$ 0.09 &  $\ldots$        & & $\ldots$         
& $\ldots$         & 1.25  \\
1158.732 &  1.39 $\pm$ 0.10 &  $\ldots$        & & $\ldots$         
& $\ldots$         & 1.47  \\
1260.927 & $\ldots$         &  5.17 $\pm$ 0.22 & & $\ldots$         
& $\leq$ 6.3       & 8.71  \\
1260.996 & $\ldots$         &  9.55 $\pm$ 0.36 & & $\ldots$         
& 6.3 $\pm$ 2.1    & 6.84  \\
1261.122 & $\ldots$         &  9.73 $\pm$ 0.30 & & 13.0 $\pm$ 1.0   
& 7.6 $\pm$ 2.1    & 10.43  \\
1329.085 & $\ldots$         & $\ldots$         & & $\ldots$         
& $\ldots$         & 12.96 \\
1329.100 & $\ldots$         & 47.83$^g$ $\pm$ 0.32 & & 42.0$^g$ $\pm$ 2.0 
& 56.0$^g$ $\pm$ 3.0 & 15.25 \\
1329.123 & $\ldots$         & $\ldots$         & & $\ldots$         
& $\ldots$         & 10.39  \\ 
J $=$ 2: &                  &                  & &                  
&                  &       \\ 
1158.019 &  2.15 $\pm$ 0.21 &  $\ldots$        & & $\ldots$         
& $\ldots$         & 4.03  \\
1158.132 &  $\ldots$        &  $\ldots$        & & $\ldots$         
& $\ldots$         & 0.44  \\
1158.397 &  0.87 $\pm$ 0.21 &  $\ldots$        & & $\ldots$         
& $\ldots$         & 0.74  \\
1261.426 & $\ldots$         &  1.95 $\pm$ 0.29 & & 3.4$^h$ $\pm$ 1.2 
& $\leq$ 6.3   & 1.45  \\
1261.552 & $\ldots$         &  4.82 $\pm$ 0.32 & & $\ldots$         
& $\leq$ 6.6      & 4.17  \\
1329.578 & $\ldots$         & 11.68$^i$ $\pm$ 0.32 & & 14.0$^i$ $\pm$ 2.0 
& 9.5$^i$ $\pm$ 3.7 & 6.60  \\
1329.601 & $\ldots$         & $\ldots$         & & $\ldots$         
& $\ldots$         & 2.34  \\ \hline
\end{tabular}

\noindent\hspace{1.0in}$^a$\ HR: High Resolution spectra, 
MR: Medium Resolution spectra.

\noindent\hspace{1.0in}$^b$\ Jenkins \& Shaya (1979).

\noindent\hspace{1.0in}$^c$\ Jenkins et al. (1983).

\noindent\hspace{1.0in}$^d$\ A: Main component B: Secondary component, 
if resolved.

\noindent\hspace{1.0in}$^e$\ Involves both the main and secondary $J=0$ 
components.

\noindent\hspace{1.0in}$^f$\ Involves a $\lambda$1158.132 (J $=$ 2) line.

\noindent\hspace{1.0in}$^g$\ Total $W_{\lambda}$ of 
$\lambda\lambda$1329.085, 1329.100, 1329.123 lines.

\noindent\hspace{1.0in}$^h$\ Total $W_{\lambda}$ of 
$\lambda\lambda$1261.426, 1261.552 lines.

\noindent\hspace{1.0in}$^i$\ Total $W_{\lambda}$ of 
$\lambda\lambda$1329.578, 1329.601 lines.
\end{table}

\newpage
\begin{table}[h]
\caption{$W_{\lambda}$ of neutral carbon lines toward 1~Sco 
\label{table:1ScoW}}
\begin{tabular}{ccccccc} \hline
                &               &                     & &                
&                      &                   \\
$\lambda$ & \multicolumn{2}{c}{Measured (m\AA)} & 
& \multicolumn{2}{c}{Other measurement} & Calculated \\ \cline{2-3} \cline{5-6}
 (\AA)    & HR$^a$           & MR$^a$           & 
&      \multicolumn{2}{c}{JJL$^b$}     &    (m\AA)    \\
                &               &                     & &                
&                      &                   \\ \hline \hline
J $=$ 0: &                  &                  & &                  & 
&       \\
1193.031A$^c$ & 10.43 $\pm$ 0.62 & $\ldots$         & 
& \multicolumn{2}{c}{$\ldots$}         &  9.35 \\
1193.031B & 11.46 $\pm$ 0.62 & $\ldots$         & 
& \multicolumn{2}{c}{$\ldots$}         & 9.56 \\
1193.996A$^d$ &  3.63 $\pm$ 0.42 & $\ldots$         & 
& \multicolumn{2}{c}{$\ldots$}        &  3.62 \\
1193.996B$^d$ &  3.02 $\pm$ 0.62 & $\ldots$         & 
& \multicolumn{2}{c}{$\ldots$}         & 3.37 \\
1260.736 & $\ldots$         & 19.73$^e$ $\pm$ 0.98 & 
& \multicolumn{2}{c}{22.0$^e$ $\pm$ 2.4} & 20.18$^e$ \\
1277.245 & $\ldots$         & 69.20$^f$ $\pm$ 1.42 & 
& \multicolumn{2}{c}{$\ldots$}         & 37.48$^e$ \\
1280.135A & $\ldots$         &  7.38 $\pm$ 0.70 & 
& \multicolumn{2}{c}{$\ldots$}         &  7.11 \\
1280.135B & $\ldots$         &  7.06 $\pm$ 0.73 & 
& \multicolumn{2}{c}{$\ldots$}         & 6.94 \\
1328.833A & 12.97 $\pm$ 0.66 & 14.50 $\pm$ 0.86 & 
& \multicolumn{2}{c}{34.0$^e$ $\pm$ 4.4} & 13.98 \\
1328.833B & 15.82 $\pm$ 0.98 & 14.50 $\pm$ 0.85 & 
& \multicolumn{2}{c}{$\ldots$}         & 15.23 \\
1560.309A & $\ldots$         & 26.72 $\pm$ 0.83 & 
& \multicolumn{2}{c}{$\ldots$}         & 21.13 \\
1560.309B & $\ldots$         & 14.15 $\pm$ 0.75 & 
& \multicolumn{2}{c}{$\ldots$}         & 25.06 \\
1656.928A & $\ldots$         & 37.89 $\pm$ 0.97 & 
& \multicolumn{2}{c}{$\ldots$}         & 28.11 \\
1656.928B & $\ldots$         & 28.58 $\pm$ 0.90 & 
& \multicolumn{2}{c}{$\ldots$}         & 37.20 \\
J $=$ 1: &                  &                  & &                  & 
&       \\
1193.009$^c$ &  2.89 $\pm$ 0.46 &  $\ldots$        & 
& \multicolumn{2}{c}{$\ldots$}         &  10.35 \\
1193.679 &  4.19 $\pm$ 0.52 &  $\ldots$        & 
& \multicolumn{2}{c}{$\ldots$}         &  4.21 \\
1194.406 &  1.46 $\pm$ 0.37 &  $\ldots$        & 
& \multicolumn{2}{c}{$\ldots$}         &  1.67 \\
1260.927 & $\ldots$         &  5.23 $\pm$ 0.94 & 
& \multicolumn{2}{c}{$\leq$ 6.6}      &  5.85 \\
1260.996 & $\ldots$         &  3.45 $\pm$ 0.78 & 
& \multicolumn{2}{c}{$\leq$ 6.6}      &  4.57 \\
1261.122 & $\ldots$         &  5.11 $\pm$ 0.81 & 
& \multicolumn{2}{c}{$\leq$ 6.0}      &  7.05 \\
1277.513$^g$ & $\ldots$         &  4.75 $\pm$ 1.15 & 
& \multicolumn{2}{c}{$\ldots$}     &  9.76 \\
1279.890 & $\ldots$         &  5.56 $\pm$ 0.74 & 
& \multicolumn{2}{c}{$\ldots$}         &  6.23 \\
1280.597 & $\ldots$         &  2.20 $\pm$ 0.90 & 
& \multicolumn{2}{c}{$\ldots$}         &  3.39 \\
1329.085 &  6.59 $\pm$ 0.52 & $\ldots$         & 
& \multicolumn{2}{c}{$\ldots$}         &  8.82 \\
1329.100 &  9.26 $\pm$ 0.56 & 26.90$^h$ $\pm$ 1.10 & 
& \multicolumn{2}{c}{38.0$^h$ $\pm$ 3.3} &  10.47 \\
1329.123 &  7.05 $\pm$ 0.63 & $\ldots$         & 
& \multicolumn{2}{c}{$\ldots$}         &  7.01 \\
1560.682 & $\ldots$         & 32.4$^i$ $\pm$ 0.98 & 
& \multicolumn{2}{c}{$\ldots$}         & 23.87 \\
1560.709 & $\ldots$         & $\ldots$         & 
& \multicolumn{2}{c}{$\ldots$}         & 12.07 \\
1656.267 & $\ldots$         & 25.89 $\pm$ 0.93 & 
& \multicolumn{2}{c}{$\ldots$}         & 25.70 \\
1657.379 & $\ldots$         & 19.09 $\pm$ 0.86 & 
& \multicolumn{2}{c}{$\ldots$}         & 19.42 \\
1657.907 & $\ldots$         & 22.75 $\pm$ 0.94 & 
& \multicolumn{2}{c}{$\ldots$}         & 22.96 \\ \hline
\end{tabular}
\end{table}
\newpage
\addtocounter{table}{-1}
\begin{table}[h]
\caption{$W_{\lambda}$ of neutral carbon lines toward 1~Sco continued}
\begin{tabular}{ccccccc} \hline
                &               &                     & &                
&                      &                   \\
$\lambda$ & \multicolumn{2}{c}{Measured (m\AA)} & 
& \multicolumn{2}{c}{Other measurement} & Calculated \\ \cline{2-3} \cline{5-6}
 (\AA)    & HR$^a$           & MR$^a$           & 
&      \multicolumn{2}{c}{JJL$^b$}     &    (m\AA)    \\
                &               &                     & 
&                &                      &                   \\ \hline \hline
J $=$ 2: &                  &                  & &                  & 
&       \\ 
1261.426 & $\ldots$         &  1.20 $\pm$ 0.94 & 
& \multicolumn{2}{c}{$\leq$ 6.3}           &  2.11 \\
1261.552 & $\ldots$         &  3.68 $\pm$ 0.68 & 
& \multicolumn{2}{c}{$\leq$ 6.3}           &  5.41 \\
1277.550$^g$ & $\ldots$         & 13.07 $\pm$ 0.87 & 
& \multicolumn{2}{c}{$\ldots$}         & 10.89 \\
1277.723 & $\ldots$         &  1.80 $\pm$ 0.89 & 
& \multicolumn{2}{c}{$\ldots$}         &  3.05 \\
1280.333 & $\ldots$         &  2.50 $\pm$ 0.90 & 
& \multicolumn{2}{c}{$\ldots$}         &  3.09 \\
1329.578 &  6.76 $\pm$ 0.55 & 10.30$^j$ $\pm$ 0.94 & 
& \multicolumn{2}{c}{$\leq$ 8.7}       &  7.91 \\
1329.601 &  4.63 $\pm$ 0.72 & $\ldots$         & 
& \multicolumn{2}{c}{$\ldots$}         &  3.30 \\
1561.340 & $\ldots$         &  3.26 $\pm$ 0.67 & 
& \multicolumn{2}{c}{$\ldots$}         &  3.78 \\
1561.438 & $\ldots$         & 13.08 $\pm$ 0.81 & 
& \multicolumn{2}{c}{$\ldots$}         & 13.42 \\
1657.008 & $\ldots$         & 15.43 $\pm$ 0.75 & 
& \multicolumn{2}{c}{$\ldots$}         & 18.04 \\
1658.121 & $\ldots$         &  8.99 $\pm$ 0.68 & 
& \multicolumn{2}{c}{$\ldots$}         &  9.91 \\ \hline
\end{tabular}

\noindent\hspace{1.0in}$^a$\ HR: High Resolution spectra, 
MR: Medium Resolution spectra.

\noindent\hspace{1.0in}$^b$\ Jenkins et al. (1983).

\noindent\hspace{1.0in}$^c$\ Blended lines.

\noindent\hspace{1.0in}$^d$\ A: Main component B: Secondary component, 
if resolved.

\noindent\hspace{1.0in}$^e$\ Involves both the main and secondary 
$J=0$ components.

\noindent\hspace{1.0in}$^f$\ Involves a second component and 
$\lambda$1277.282.

\noindent\hspace{1.0in}$^g$\ Blended lines.

\noindent\hspace{1.0in}$^h$\ Total $W_{\lambda}$ of 
$\lambda\lambda$1329.085, 1329.100, 1329.123 lines.

\noindent\hspace{1.0in}$^i$\ Total $W_{\lambda}$ of 
$\lambda\lambda$1560.682, 1560.709 lines.

\noindent\hspace{1.0in}$^j$\ Total $W_{\lambda}$ of 
$\lambda\lambda$1329.578, 1329.601 lines.
\end{table}

\newpage
\begin{table}
\caption{$W_{\lambda}$ of neutral carbon lines toward $\delta$~Sco 
\label{table:dScoW}}  
\begin{tabular}{ccccccc} \hline
                &                &           & &            &          
&    \\
$\lambda$ & \multicolumn{2}{c}{Measured
(m\AA)} & & \multicolumn{2}{c}{Other measurement} & Calculated \\ 
\cline{2-3} \cline{5-6}
 (\AA)    & HR$^a$           & MR$^a$           & 
&      \multicolumn{2}{c}{JS$^b$}     &   (m\AA)     \\
                &                &           & &            &          
&    \\ \hline \hline 
J $=$ 0: &                  &                  & &                  & 
&       \\
1157.910A$^c$ & 23.06 $\pm$ 0.28 & $\ldots$         & 
& \multicolumn{2}{c}{$\ldots$}         & 20.36 \\
1157.910B$^c$ &  2.11 $\pm$ 0.20 & $\ldots$         & 
& \multicolumn{2}{c}{$\ldots$}         & 4.87 \\
1158.324A & 10.84 $\pm$ 0.23 & $\ldots$         & 
& \multicolumn{2}{c}{$\ldots$}         & 10.78 \\
1158.324B &  1.44 $\pm$ 0.25 & $\ldots$         & 
& \multicolumn{2}{c}{$\ldots$}         & 1.89 \\
1260.736A & 21.46 $\pm$ 0.43 & 28.53$^d$ $\pm$ 0.22 & 
& \multicolumn{2}{c}{32.0$^d$ $\pm$ 3.0} & 22.72 \\
1260.736B &  6.03 $\pm$ 0.36 & $\ldots$             & 
& \multicolumn{2}{c}{$\ldots$}         & 5.56 \\
J $=$ 1: &                  &                  & 
&                  & &       \\
1157.770 &  5.06 $\pm$ 0.21 &  $\ldots$        & 
& \multicolumn{2}{c}{$\ldots$}         &  5.38 \\
1158.035$^e$ &  7.98 $\pm$ 0.21 &  $\ldots$        & 
& \multicolumn{2}{c}{$\ldots$}         &  8.51 \\
1158.130$^f$ &  4.05 $\pm$ 0.22 &  $\ldots$        & 
& \multicolumn{2}{c}{$\ldots$}         &  4.24 \\
1158.674 &  1.37 $\pm$ 0.14 &  $\ldots$        & 
& \multicolumn{2}{c}{$\ldots$}         &  1.32 \\
1158.732 &  1.59 $\pm$ 0.13 &  $\ldots$        & 
& \multicolumn{2}{c}{$\ldots$}         &  1.53 \\
1260.927 &  7.63 $\pm$ 0.30 &  8.00 $\pm$ 0.20 & 
& \multicolumn{2}{c}{$\ldots$}         &  7.99 \\
1260.996 &  6.54 $\pm$ 0.31 &  6.50 $\pm$ 0.19 & 
& \multicolumn{2}{c}{7.2 $\pm$ 1.0}  &  6.49 \\
1261.122 &  8.15 $\pm$ 0.31 &  8.80 $\pm$ 0.21 & 
& \multicolumn{2}{c}{9.7 $\pm$ 2.0}  &  9.27 \\
J $=$ 2: &                  &                  & &                  & 
&       \\ 
1158.019$^e$ &  4.06 $\pm$ 0.26 &  $\ldots$        & 
& \multicolumn{2}{c}{$\ldots$}         &  3.94 \\
1158.132$^f$ &  0.43 $\pm$ 0.22 &  $\ldots$        & 
& \multicolumn{2}{c}{$\ldots$}         &  0.49 \\
1158.397 &  0.81 $\pm$ 0.21 &  $\ldots$        & 
& \multicolumn{2}{c}{$\ldots$}         &  0.81 \\
1261.426 &  1.61 $\pm$ 0.30 &  2.74 $\pm$ 0.26 & 
& \multicolumn{2}{c}{$\ldots$}         &  1.55 \\
1261.552 &  3.87 $\pm$ 0.35 &  4.95 $\pm$ 0.21 & 
& \multicolumn{2}{c}{4.6 $\pm$ 1.4}  &  4.09 \\ \hline
\end{tabular}

\noindent\hspace{1.0in}$^a$\ HR: High Resolution spectra, 
MR: Medium Resolution spectra.

\noindent\hspace{1.0in}$^b$\ Jenkins \& Shaya (1979).

\noindent\hspace{1.0in}$^c$\ A: Main component B: Secondary component, 
if resolved.

\noindent\hspace{1.0in}$^d$\ Involves both the main and secondary 
$J=0$ components.

\noindent\hspace{1.0in}$^e$\ Blended lines.

\noindent\hspace{1.0in}$^f$\ Blended lines.
\end{table}

\newpage
\begin{table}[h]
\caption{Derived parameter values \label{table:fit}} 
\begin{tabular}{lccccc} \hline
              & Doppler shift        & & Column density                & 
& $b$-value       \\
              &     (km s$^{-1}$)    & &          (cm$^{-2}$)          & 
&  (km s$^{-1}$)  \\ \hline
$\lambda$~Ori &                      & &                               & 
&                 \\
Component I.  &                      & &                               & 
&                 \\
J= 0          &    -0.11             & &  1.33 $\times$ 10$^{14}$       & 
&        3.02     \\
J= 1          &     0.28             & &  5.52 $\times$ 10$^{13}$       & 
&        2.35     \\
J= 2          &    -1.05             & &  9.24 $\times$ 10$^{12}$       & 
&        2.50     \\
Component II. &                      & &                               & 
&                 \\
J= 0          &    -12.11            & &  9.55 $\times$ 10$^{12}$       & 
&        3.00     \\
J= 1          &    -12.11            & &  1.72 $\times$ 10$^{12}$       & 
&        3.00     \\
J= 2          &    -12.11            & &  1.43 $\times$ 10$^{12}$       & 
&        3.00     \\
              &                      & &                               & 
&                 \\
1~Sco         &                      & &                               & 
&                 \\
Component I.  &                      & &                               & 
&                 \\
J= 0          &     0.61             & &  2.49 $\times$ 10$^{13}$       & 
&  1.82           \\
J= 1          &     0.49             & &  3.31 $\times$ 10$^{13}$       & 
&  1.48           \\
J= 2          &     0.01             & &  1.61 $\times$ 10$^{13}$       & 
&  1.23           \\
Component II. &                      & &                               & 
&                 \\
J= 0          &    -7.13             & &  2.21 $\times$ 10$^{13}$       & 
&  3.05           \\
J= 1          &    -7.13             & &  4.32 $\times$ 10$^{12}$       & 
&  2.49           \\
J= 2          &    -7.13             & &  3.96 $\times$ 10$^{11}$       & 
&  2.50           \\
              &                      & &                               & 
&                 \\
$\delta$~Sco  &                      & &                               & 
&                 \\
Component I.  &                      & &                               & 
&                 \\
J= 0          &    -0.01             & &  8.87 $\times$ 10$^{13}$       & 
&        2.32     \\
J= 1          &     0.03             & &  5.78 $\times$ 10$^{13}$       & 
&        1.19     \\
J= 2          &     0.14             & &  1.19 $\times$ 10$^{13}$       & 
&        1.16     \\
Component II. &                      & &                               & 
&                 \\
J= 0          &    -4.76             & &  1.25 $\times$ 10$^{13}$       & 
&        1.41     \\
J= 1          &    -4.76             & &  3.82 $\times$ 10$^{12}$       & 
&        1.50     \\
J= 2          &    $\ldots$          & & $\leq$ 6.72 $\times$ 10$^{11}$ & 
&   $\ldots$      \\ \hline
\end{tabular}
\end{table}

\newpage
\begin{table}[h]
\caption{Revised neutral carbon oscillator strengths \label{table:Newfvalues}}
\begin{tabular}{cccccccccc} \hline
$\lambda$ & Lower & Upper & $f$(ZsF)$^a$ & $f$(ZsFC)$^b$ &  $f$(M)$^c$ 
& $f$(H)$^d$ & $f$(WFD)$^e$ & $f$(W)$^f$ & $\frac{f({\rm M})}{f({\rm ZsF})}$ \\
\cline{4-9}
(\AA) & State & State & \multicolumn{6}{c}{$\times\ 10^{-3}$} & \\ \hline \hline
1194.406 & $^3P_1$ & 5$s$ $^3P^{\rm o}_0$ & 3.70(1.01)$^g$ & $\cdots$ 
& 3.14 & 3.32 & 3.19 & $\cdots$ & 0.85 \\
1193.996 & $^3P_0$ & 5$s$ $^3P^{\rm o}_1$ & 12.8(2.9) & 14.2 
& 9.41 & 13.3 & 12.8 & $\cdots$ & 0.66 \\
1193.679 & $^3P_1$ & 5$s$ $^3P^{\rm o}_2$ & 10.1(1.6) & 9.00 
& 3.92 & 10.6 & 10.2 & 6.44 & 0.39 \\
1193.031 & $^3P_0$ & 4$d$ $^3D^{\rm o}_1$ & 41.1(4.8) & 54.8 
& 44.5 & 41.0 & 45.1 & $\cdots$ & 1.08 \\
1193.009 & $^3P_1$ & 4$d$ $^3D^{\rm o}_2$ & 30.9(5.8) & 46.8 
& 33.4 & 26.4 & 29.1 & $\ldots$ & 1.08 \\
1158.732 & $^3P_1$ & 5$d$ $^3F^{\rm o}_2$ & 2.23(0.29) & $\cdots$ 
& 1.88 & $\cdots$ & $\cdots$ & $\cdots$ & 0.84 \\
1158.674 & $^3P_1$ & 6$s$ $^3P^{\rm o}_0$ & 1.90(0.27) & $\cdots$ 
& 1.14 & $\cdots$ & 1.86 & $\cdots$ & 0.60 \\
1158.397 & $^3P_2$ & 6$s$ $^3P^{\rm o}_2$ & 5.91(1.64) & $\cdots$ 
& 2.57 & $\cdots$ & 4.18 & $\cdots$ & 0.43 \\
1158.324 & $^3P_0$ & 6$s$ $^3P^{\rm o}_1$ & 13.7(2.8) & 11.1 
& 3.42 & $\cdots$ & 5.57 & $\cdots$ & 0.25 \\
1158.132 & $^3P_2$ & 5$d$ $^3D^{\rm o}_2$ & 3.53(1.84) & $\cdots$ 
& 3.27 & $\cdots$ & 3.66 & $\cdots$ & 0.93 \\
1158.130 & $^3P_1$ & 5$d$ $^3D^{\rm o}_1$ & 6.97(0.79) & 1.90 
& 5.44 & $\cdots$ & 6.09 & $\cdots$ & 0.78 \\
1158.035 & $^3P_1$ & 6$s$ $^3P^{\rm o}_2$ & 17.8(1.8) & $\cdots$ 
& 1.43 & $\cdots$ & 2.32 & $\cdots$ & 0.08 \\
1158.019 & $^3P_2$ & 5$d$ $^3D^{\rm o}_3$ & 34.0(4.0) & 6.41 
& 18.3 & $\cdots$ & 20.5 & $\cdots$ & 0.54 \\
1157.910 & $^3P_0$ & 5$d$ $^3D^{\rm o}_1$ & 40.5(5.6) & 15.6 
& 21.8 & $\cdots$ & 24.4 & $\cdots$ & 0.54 \\
1157.770 & $^3P_1$ & 5$d$ $^3D^{\rm o}_2$ & 9.36(1.01) & 5.71 
& 16.3 & $\cdots$ & 18.3 & 5.79 & 1.74 \\ \hline
\end{tabular}

\noindent\hspace{0.5in}$^a$\ Present compilation.

\noindent\hspace{0.5in}$^b$\ Zsarg\'{o} et al. (1997), with corrections to 
$f$-values for J=0 lines as noted in the text.

\noindent\hspace{0.5in}$^c$\ Morton (1991).

\noindent\hspace{0.5in}$^d$\ Hibbert et al. (1993).

\noindent\hspace{0.5in}$^e$\ Wiese et al. (1996).

\noindent\hspace{0.5in}$^f$\ Welty et al. (1999); only suggested changes to 
$f$-values of Wiese et al. and Morton 

\noindent\hspace{0.5in}~     are listed.

\noindent\hspace{0.5in}$^g$\ 1-$\sigma$ uncertainties given in parentheses.
\end{table}

\end{document}